\documentclass[fleqn,twoside]{article}
\usepackage{espcrc2}

\usepackage{graphicx}
\usepackage[figuresright]{rotating}

\newcommand{\AmS}{{\protect\the\textfont2
  A\kern-.1667em\lower.5ex\hbox{M}\kern-.125emS}}
\def\ltsim{\hbox{\raise 2pt \hbox {$<$} \kern-1.1em \lower 4pt \hbox {$\sim$}}}
\def\ltapprox{\hbox{\raise 2pt \hbox {$<$} \kern-1.1em \lower 5pt \hbox 
{$\approx$}}}
\def\gtsim{\hbox{\raise 2pt \hbox {$>$} \kern-1.1em \lower 4pt \hbox {$\sim$}}}
\def\gtapprox{\hbox{\raise 2pt \hbox {$>$} \kern-1.1em \lower 5pt \hbox 
{$\approx$}}}
\def\arcsec{$^{\prime\prime}$}

\title{Magnetic fields in Clusters of Galaxies}

\author{L. Feretti\address{Istituto di Radioastronomia CNR/INAF\\
        Bologna, Italy, {\it lferetti@ira.cnr.it}},
        M. Johnston-Hollitt\address{Leiden Observatory\\
        Leiden, The Netherlands, {\it johnston@strw.leidenuniv.nl}}}
       
\begin{document}

\begin{abstract}
An important area of study of cosmic magnetic fields is on the largest
scales, those of clusters of galaxies.  In the last decade it has 
become clear that the intra-cluster medium (ICM) in clusters of
galaxies is magnetized and that magnetic fields play a critical role
in the cluster formation and evolution. The observational evidence for
the existence of cluster magnetic fields is obtained by the diffuse
cluster-wide synchrotron radio emission and from rotation
measure (RM) studies of extragalactic radio sources located within or
behind the clusters.  A significant breakthrough in the knowledge of
the cluster magnetic fields will be reached through the SKA, owing to
its capabilities, in particular the deep sensitivity and the polarization
purity.
\vspace{1pc}
\end{abstract}

\maketitle

\section{INTRODUCTION}

Unlike electromagnetic radiation from astrophysical sources, distant
magnetic fields are difficult to detect.  Nonetheless, recent
measurements have begun to reveal that such fields exist at
significant strengths, and on surprisingly large scales, in the
extragalactic universe.  Observations have shown that magnetic fields
are ubiquitous in cluster atmospheres, playing a critical role in the
cluster formation and evolution, in determining the energy balance in
cluster gas through their effect on heat conduction, and in some
cases, perhaps even becoming dynamically important.

Our knowledge of the magnetic field properties in galaxy clusters and
of how they relate to other cluster properties is limited by the
sensitivity and resolution of current instruments.  In particular,
several questions are still unanswered: are the fields filamentary,
what are the coherence scales, to what extent do the thermal and
non-thermal plasmas mix in cluster atmospheres, what is the radial
dependence of the field strength, how does the field depend on cluster
parameters such as the gas temperature, metallicity, mass,
substructure and density profile, how do the fields evolve with cosmic
time, how do the fields extend and finally how were these fields
generated?

Magnetic fields associated with the intracluster medium (ICM) in
galaxy clusters are investigated in the radio band through studies of
the rotation measure and of the synchrotron emission of both
individual radio galaxies and radio halos and relics. Other techniques
include X-ray studies of the inverse Compton emission and of cold
fronts, and hydrodynamic simulations.  The studies in the radio band
are, however, the most relevant and provide the most accurate field
estimates.  The SKA's high sensitivity, high resolution,
multifrequency capability, and polarization purity will be crucial to
these studies.

\section{CURRENT RESULTS FROM RM STUDIES}

Cluster surveys of the Faraday rotation measures of polarized radio
sources both within and behind clusters provide an important probe of
the existence of intracluster magnetic fields, as the radio waves
traversing the magnetized intracluster medium show depolarization and
rotation of the polarization position angle as a function of
wavelength.

\begin{figure}
\includegraphics[height=16pc]{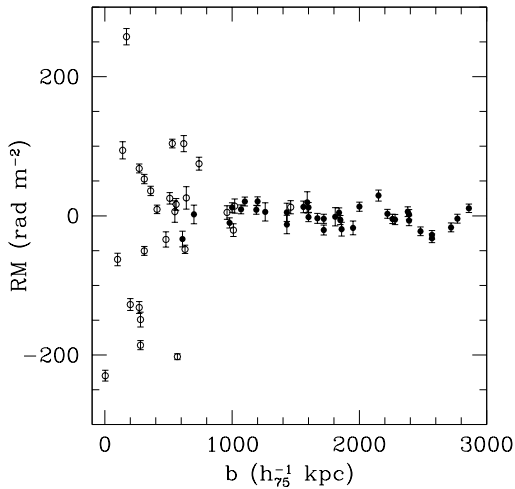}
\caption{
Galaxy-corrected rotation measure plotted as a
function of source impact parameter in kiloparsecs for the sample of
sources from \cite{Clar01}. Open dots refer to cluster sources,
closed dots to control sources.
\label{rmstat}
}
\end{figure}

The RM values derived from multifrequency polarimetric observations of
background or embedded cluster sources are of the order of tens to
thousands rad m$^{-2}$. These have to be combined with measurements of the
thermal gas density, $n_{\rm e}$, to estimate the cluster magnetic field
along the line of sight.  The observing strategy to derive information
on the magnetic field intensity and structure is twofold: i) obtain
the average value of the RM of sources located at different impact
parameters of the cluster, ii) derive maps of the RM of extended radio
sources, to evaluate the $\sigma$ of the RM distribution.

Part of the difficulty of investigating cluster magnetic fields
through Faraday rotation is that at present such a study may only be
undertaken statistically over a large number of clusters. This is due,
in part, to the lack of available RMs which are limited by current
instrument sensitivity and by the vectorial nature of the RMs
themselves.  Since every measured RM is the vectorial addition of all
contributing Faraday screens along the line of sight it is impossible
to disentangle the cluster rotation measure components from either
internal rotation in the source, or a Galactic rotation measure
component without a sufficient number of RMs available.

Nevertheless, studies on both statistical samples and individual
clusters have been carried out (see the review \cite{CT02} and
references therein).  Kim et al.\cite{Kim91} analyzed the RM of 53
radio sources in and behind clusters and 99 sources in a control
sample.  This study contains the largest cluster sample to date. It
demonstrated that $\mu$G level fields are widespread in the ICM,
regardless of whether there is the direct evidence for the existence
of magnetic fields from the presence of diffuse radio emission.  In a
more recent statistical study, Clarke et al.\cite{Clar01} analyzed the
RMs for a representative sample of 27 cluster sources, plus a control
sample, and found a statistically significant broadening of the RM
distribution in the cluster sample, and a clear increase in the width
of the RM distribution toward smaller impact parameter (see
Fig. \ref{rmstat}).  They derived that the ICM is permeated with a
high filling factor by magnetic fields at levels of 4 - 8 $\mu$G and
with a correlation length of $\sim$15 kpc, up to $\sim$0.75 Mpc from
the cluster center.

The first detailed studies of RM have been performed on cooling core
clusters, owing to the extremely high RMs of the powerful radio
galaxies at their centers (e.g., Hydra A \cite{TP93} \& 3C295
\cite{Allen01}).  High values of the magnetic fields, up to tens of
$\mu$G, have been obtained, but they only refer to the innermost
cluster regions.

\begin{figure}
\includegraphics[height=16pc]{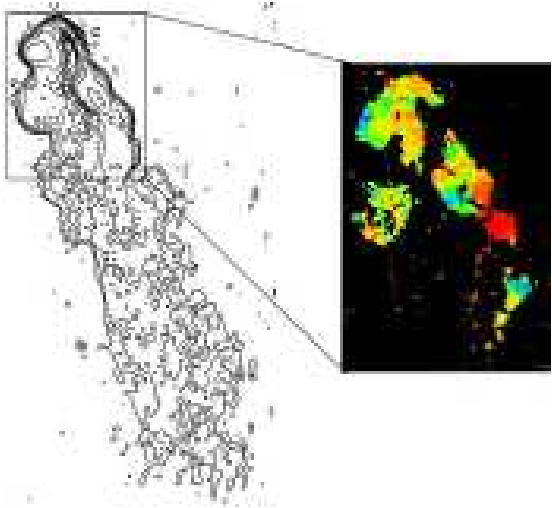}
\caption{VLA contour plot of the tailed radio galaxy
 0053-015 in A119 at 1.4 GHz (left), and RM image (right).
The RM information has only been derived for the
brightest source region \cite{Fer99}.
\label{rmsource}
}
\end{figure}

Studies on larger areas of clusters have been carried out e.g. for
Coma \cite{Fer95}, A119 \cite{Fer99}, A514 \cite{Gov01} and 3C129
\cite{Tayl01}. However, because of the limited sensitivity of current
instruments, reliable maps of the RM can be obtained only for the
strongest sources (total flux \gtsim 50 mJy) and only in the regions
of high radio surface brightness (see e.g. Fig. \ref{rmsource}).  The
number of targeted sources per cluster is on average 1-2, with a
maximum of 3-5 in a few clusters.

Overall, the data are consistent with cluster atmospheres containing
$\mu$G fields, with perhaps an order of magnitude scatter in field
strength between clusters, or within a given cluster, and with extreme
field values in cluster cooling cores.  The RM distribution is
generally patchy, indicating that large scale magnetic fields are not
regularly ordered on cluster scales, but have coherence scales between
1 and 10 kpc.  The estimates of the magnetic field strength crucially
depend on the magnetic field structure and geometry, thus accurate
maps of RM are needed.  We note that in the Coma cluster the presence
of a weaker field component of 0.1-0.2 $\mu$G, ordered on a scale of
about one cluster core radius, was inferred in addition to the
stronger tangled magnetic field component \cite{Fer95}.

A caveat in the interpretation of RM data is the possible existence of
local RM enhancements, produced by the compression of the ICM fields
by a radio galaxy. In this case the RM would not be indicative of the
cluster magnetic field, leading to overestimates of the ICM magnetic
field strength \cite{RB03}.  This difficulty can be overcome by
sensitive observations at very high resolution, which cannot be
obtained with the presently available instruments.

The observations are often interpreted in terms of the simplest
possible model, i.e. in this case a constant field throughout the
whole cluster.  However, a decline with radius is expected if the
intensity of the magnetic field results from the compression of the
thermal plasma during the cluster gravitational collapse.  According
to this model, a correlation between observable parameters, the RM and
the X-ray surface brightness, is expected to reflect the correlation
between the physical quantities, magnetic field and gas density.  The
application of this approach has been possible so far only in A119,
giving the radial profile of the magnetic field as $B \propto
n_e^{0.9}$ \cite{Dol01} in this cluster. In addition, Beck et 
al.\cite{Bec03} pointed out that field estimates derived from RM may be
too large in the case of a turbulent medium where small-scale
fluctuations in the magnetic field and the electron density are highly
correlated.

New generation instruments are rather promising and establish a clear
connection between radio astronomical techniques and the improvement in
the knowledge of the X-ray sky. There are various satellite missions
which will map the X-ray sky at low energies in the next years. These
will provide a more precise knowledge of the X-ray surface brightness
of clusters, i.e. of their thermal gas density, allowing
a more accurate and correct interpretation of  the sensitive RM 
measurements. The accurate experimental determination of large scale
magnetic fields in the intracluster medium will thus be possible.

\section{DIFFUSE SYNCHROTRON \\ EMISSION}

\begin{figure}
\includegraphics[height=14pc]{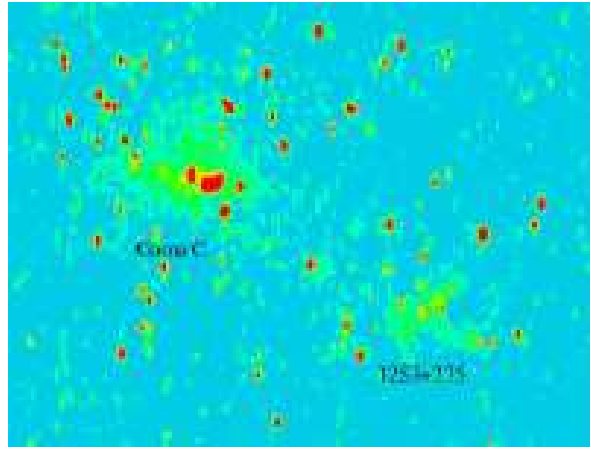}
\caption{Radio image of the Coma cluster region at 90 cm,
with angular resolution of 55$^{\prime\prime}$ $\times$
125$^{\prime\prime}$ (HPBW, RA $\times$ DEC), showing
the halo source Coma C at the cluster center
and the relic source 1253+275 at the cluster periphery.
\label{comaclus}
}
\end{figure}

The presence of magnetic fields in clusters is directly demonstrated
by the existence of the radio halos and relics, i.e.  diffuse
cluster-wide synchrotron radio sources, as revealed in Coma
(Fig. \ref{comaclus})
and some other clusters \cite{GF02}. Under the assumption that
the energy density within radio sources is minimum (equipartition
condition), magnetic field values in the range 0.1-1 $\mu$G are
derived for the radio emitting regions, i.e. on scales as large as
$\sim 1$ Mpc. 
These calculations typically assume equal energy in relativistic 
protons and electrons, a  magnetic field entirely filling the radio source
volume,  a low frequency cut-off of 10 MHz, and a high frequency 
cut-off of 10 GHz. The magnetic field values derived in this
way are consistent with those suggested from
the recent detection of Inverse Compton hard X-ray emission in
clusters with halos or relics \cite{RFF03}.

The number of clusters presently known to host halos and relics is
around 50, i.e. $\sim$ 10\% of rich clusters. Indeed,
the typically low surface brightness of cluster radio halos and their
steep spectrum makes it difficult to image them accurately with the
current resources.  Further, at lower resolution, where beam averaging
enhances the detectability of extended radio emission, true diffuse
emission is sometimes difficult to distinguish from a blend of weak,
discrete radio sources. 

The observations of clusters with the SKA will allow a dramatic
improvement of the knowledge of halos and relics. It will be possible
to detect new halos and relics, and study these sources in great
detail (see detailed discussion Feretti, Burigana, \& Ensslin, this
volume). In particular, polarimetric studies will be of crucial
importance, to give direct information on the magnetic field
orientation, degree of ordering, and overall structure.

The detection of synchrotron radiation at the lowest possible levels
will allow the measurement of magnetic fields in even more rarefied
regions of the intergalactic space, and investigations of the
relation between the formation of magnetic fields and the formation
of the large scale structure in the universe.

\begin{table*}
   \caption{Magnetic field estimates derived from various methods for the galaxy cluster A3667. Column 1 gives the method used to estimate the field strength,
Column 2 the value of the magnetic field in $\mu$G, Column 3 describes the
location in the cluster at which this estimation is made and Column 4 gives the
reference.}
\medskip\medskip
   \begin{center}

 \begin{tabular}{l|c|c|c}
        \hline

  Method & Field Estimate ($\mu$G) & Location in the cluster& Reference \\
         \hline \hline
  Inverse Compton  & $\geq$ 0.4   & cluster core& \cite{RFF01}\\
  Kelvin-Helmholtz & 7-16         & along the cold front& \cite{Vik01}\\
  Faraday Rotation & 1-2          & cluster core& \cite{mjh03}\\
  Faraday Rotation & 3-5          & NW radio emission region& \cite{mjh03}\\
  Equipartition    & 1.5-2.5      & NW radio emission region& \cite{mjh03}\\
         \hline\hline

 \end{tabular}
 \end{center}
 \label{tab:est}
\end{table*}

\section{RECONCILING MAGNETIC FIELD VALUES}

The cluster magnetic field values obtained from RM arguments are about
an order of magnitude higher than the estimates, typically of 0.2 to 1
$\mu$G, derived from both the synchrotron diffuse radio emission
\cite{GF02} and the inverse Compton hard X-ray emission \cite{RFF03}.
The discrepancy can be alleviated by taking into account that i) the
values deduced from radio synchrotron emission and from inverse
Compton refer to averages over large volumes, whereas the RM estimates
give a weighted average of the field along the line of sight; ii) the
magnetic field intensity is likely to decline with the distance from
the cluster center; iii) the magnetic field may show complex
structure, as filamentation and/or substructure with a range of
coherence scales.  Therefore, the RM data should be interpreted using
realistic models of the cluster magnetic fields, as shown by a recent
investigation performed using a numerical approach \cite{GM03,Mur04}.

Additionally, evidence suggests that the magnetic field strength will
vary depending on the dynamical history and location within the
cluster. A striking example of the variation of magnetic field
strength estimates for various methods and in various locations
throughout the cluster is given in Table 1 for the post merger cluster
A3667 (at z = 0.055). The results of each estimation, are consistent
with a typical 1--2 $\mu$G field, tangled on scales of 10 to 100 kpc,
pervading the cluster's central region. This field has been further
enhanced in the region of the observed central X-ray cold front to a
level of 7--16 $\mu$G \cite{Vik01} and to around 3--5 $\mu$G in the
region of the Mpc-scaled radio relic in the northern part of the
cluster \cite{mjh03}. As the relic emission is currently thought to be
the result of a cluster merger it is likely that the central field
would be compressed and elevated by a factor of 3--4 in the region of
the shock accelerated relic.

As shown in Table 1 the observational data needed to obtain
sufficiently detailed information about the cluster magnetic field
strength and structure can only be achieved by a new generation
instrument.

\section{RELEVANCE OF STUDIES WITH SKA}

Although several interesting results have been
obtained in recent years about the cluster
magnetic fields, such studies are still limited
to works on a few clusters, and on a few radio sources,
as reported in the previous sections. Magnetic field
strengths of the order of $\sim$ 1 $\mu$G are found to be
common in clusters. However, estimates obtained with different
approaches may differ (Sec. 4), thus detailed information
on the cluster field is still needed.

The main goals to pursue with SKA are the following:

$\bullet$ 
obtain the RM of large samples of radio sources
within or behind clusters;

$\bullet$ 
derive detailed RM maps at high resolution,  
to resolve small scale features in the
foreground screens, in particular those
due to turbulence and other local effects;

$\bullet$ 
distinguish the external Faraday rotation measure from that
arising internally in a radio source, in order to 
get careful information on the cluster magnetic field
(and on the radio source too);

$\bullet$ 
investigate the ICM magnetic field structure, the 
existence of components with different coherence lengths
(power spectrum), the magnetic field filling factor;

$\bullet$ 
analyze the correlation between the magnetic field intensity 
and the gas density, get information on the magnetic field
profile,

$\bullet$ 
test models of magnetic field formation from the study of distant
objects and the effect of density inhomogeneities on their Faraday 
RM.

The SKA specifications to these aims are

$\bullet$  
a frequency range  1 - 10 GHz with a large number of
channels\cite{Deb96}, 
to get reliable information on the RM, solving the
ambiguity related to its computation and disentangling
various contributions; 

$\bullet$ 
an angular resolution  of 0.5\arcsec~ at 1.4 GHz 
to allow structures of 1 kpc at z \ltsim 0.5 to be resolved;
a resolution of 0.1\arcsec~ is needed for the distant
clusters.
          
$\bullet$    
polarization purity of -40 dB at the field center
to allow measurement of the polarization parameters
for submJy sources.

The All Sky RM survey (SKA Key Project on Magnetic Fields
\cite{BG04}), will provide the RM for \gtsim$10^7$ compact polarized
extragalactic sources, expanding the sample of RM measurements by five
orders of magnitude over current data sets.  When combined with
redshift and X-ray information from future instruments, such a
data-base will allow a statistical analysis of the RM of sources
within or behind clusters, leading to a great improvement of the
knowledge of the strength of the magnetic field in clusters.  For
example in the case of A3667 illustrated above only 3 reliable RMs
were obtained after over 100 hours of observations on present day
instruments. With the All Sky RM Survey, it will be possible to obtain
about 1000 RM measurements through lines of sight in this cluster, and
generally in clusters at similar distances.  Additionally, we will
detect at least 20 RMs through clusters up to redshifts of 3.5 giving
the first opportunity to perform detailed analysis of the evolution of
cluster magnetic fields.

Deep multifrequency surveys, targeting individual clusters, will allow
the investigation of the intensity and structure of the ICM magnetic
fields.  Low surface brightness radio features such as relics and
halos should be detected in their thousands \cite{ER} allowing us to
explore the role of dynamically important magnetic fields in merging
clusters and providing vital clues to the origin of cluster magnetic
fields.

Using the described techniques, magnetic fields can also be observed and
studied in the jets and lobes of radio galaxies. The largely improved
sensitivity and resolution of SKA will allow the study of faint
objects and of the low brightness components of extended radio
sources, the distinction of local features, the discrimination between
internal and external Faraday dispersion, the connection between
magnetic fields within the radio galaxies and the cosmological
magnetic fields.

\medskip
{\bf ACKNOWLEDGMENTS}

\noindent
We are grateful to R. Beck and G. Giovannini for interesting
discussions and helpful suggestions.

\end{document}